\documentclass[11pt]{article}
\usepackage[utf8]{inputenc}
\usepackage[width=6.5in]{geometry}
\usepackage{amssymb,amsfonts,amsmath,enumerate,mathrsfs}
\usepackage{booktabs}
\usepackage{float,adjustbox,tabularx,tabulary}
\usepackage{multirow,tikz}
\usepackage{subcaption}
\usepackage[linguistics]{forest}
\usepackage{titlesec}
\usepackage{dcolumn,rotating, graphicx}

\usepackage{graphicx}
\usepackage{amsmath}
\usepackage{algorithm}
\usepackage{algorithmic}

\usepackage[utf8]{inputenc}
\usepackage{bm,booktabs}
\usepackage{amsmath}
\usepackage{thumbpdf}

\usepackage[square,sort,comma,numbers]{natbib}
\usepackage{appendix}

\newcolumntype{L}{>{\raggedright\arraybackslash}p}
\newcolumntype{R}{>{\raggedleft\arraybackslash}p}
\newcolumntype{C}{>{\centering\arraybackslash}p}

\usepackage{threeparttable}

\begin{document}
	
\begin{titlepage}

\title{\vspace{-2.0cm} Disentangling shock diffusion on complex networks: Identification through graph planarity
\thanks{We are grateful to Michele Tumminello for helping us with the code for constructing partial correlation planar graphs analyzed in the paper. We would also like to thank the HPC lab at IIM Ahmedabad and NVIDIA corporation for the donation of a GPU, which helped us to carry out the computations presented in the paper. This research was partially supported by 
institute grant, IIM Ahmedabad, ESRC Network Plus project `Rebuilding macroeconomics’ and Economic and Political Science
Research Council (EPSRC) grant EP/P031730/1. All data have ben obtained from Thomson Reuters Eikon database accessed through the Vikram Sarabhai Library, IIM Ahmedabad. All remaining errors are ours.}}
\author{Sudarshan Kumar\thanks{{\it email}: sudarshank@iima.ac.in. Finance \& accounting area, Indian Institute of Management, Vastrapur, Ahmedabad, Gujarat-380015, India.}
\and Tiziana Di Matteo\thanks{{\it email}: tiziana.di\_matteo@kcl.ac.uk, Department of Mathematics, King’s College London, The Strand, London, WC2R 2LS, UK; Department of Computer Science, University College London, Gower Street, London, WC1E 6BT, UK; Complexity Science Hub Vienna, Josefstaedter Strasse 39, A 1080 Vienna, Austria.}
\and Anindya S. Chakrabarti\thanks{(corresponding author) {\it email}: anindyac@iima.ac.in. Economics area, Indian Institute of Management, Vastrapur, Ahmedabad, Gujarat-380015, India.}
}

\date{\today}
\maketitle

\begin{abstract}
\noindent Large scale networks delineating collective dynamics often exhibit cascading failures across nodes leading to a system-wide collapse. Prominent examples of such phenomena would include collapse on financial and economic networks. Intertwined nature of the dynamics of nodes in such network makes it difficult to disentangle the source and destination of a shock that percolates through the network, a property known as {\it reflexivity}. In this article, a novel methodology is proposed which combines 
{\it vector autoregression} model with an unique identification restrictions obtained from the topological structure of the network to uniquely characterize cascades. In particular, we show that {\it planarity} of the network allows us to statistically estimate a dynamical process
consistent with the observed network and thereby uniquely identify a path for shock propagation from any chosen epicenter
to all other nodes in the network. We analyze the distress propagation mechanism in closed loops giving rise to a detailed picture of the effect of feedback loops in transmitting shocks. We show usefulness and applications of the algorithm in two networks with dynamics at different time-scales: worldwide GDP growth network and stock network. In both cases, we observe that the model predicts the impact of the shocks emanating from the US would be concentrated within the cluster of developed countries and the developing countries show very muted response, which is consistent with empirical observations over the past decade.
\end{abstract}

\vskip .51 cm




\end{titlepage}

\section{Introduction}

In the post-global financial crisis period, networks have become standard descriptions of the increasingly interlinked world (\cite{Acemoglu_16}).
A significant stream of research in network theory attempts to disentangle the mechanics of distress propagation from a given epicenter to the entire network. Much of the present literature focuses on financial networks due to its immediate importance and applicability (see e.g. \cite{battiston2007credit}, \cite{battiston2012default}, \cite{wang2017extreme}, \cite{Diebold_book_15}). However, similar mathematical mechanisms are studied in context of many other types of complex networks as well; e.g. in epidemics (\cite{zhan2018coupling}) and social contagion (\cite{chang2018co}).

A critical problem in this literature is to disentangle individual impacts of nodes on the network behavior from the collective dynamics displayed by the network as a whole. In case of network models of {\it social interactions}, this problem is referred to as the {\it reflection} problem \cite{manski2000economic}. The essential problems are threefold: one, behavior across nodes can be interdependent making it difficult to disentangle cause and effects; two, all nodes might react to similar exogenous shocks; three, all nodes may possess very similar characteristics leading to correlated behavior. The same set of problems would feature in case of a network characterizing interdependent dynamical systems. 
Therefore, the extant literature has focused on the function of individual nodes and edges in the mechanism of shock propagation in diverse types of networks ranging from economic and financial (e.g. \cite{betz2016systemic}, \cite{hautsch2014financial}) to bio-physical networks (e.g. \cite{jalili2017information}, \cite{zhan2019information}).

In this article, we present an algorithmic approach 
to analyze distress propagation in statistically estimated networks characterizing 
an interdependent dynamical system. In order to perform the analysis, we propose a two-step mechanism. In step one,
we first estimate a statistical model of the interconnected dynamical system in the form of a {\it vector autoregression}
model (VAR henceforth). Such a model captures cross-dependence of the dynamics of the nodes as well as the temporal dimension of evolution of the nodes. Here, each node represents one time series. In the second step,
we utilize the topological properties of the network to uniquely pin down the paths of  distress propagation.
In particular, we propose that a filter based on planarity of the constructed network provides the most useful information about the linkages to be preserved for analyzing distress propagation. Below we elaborate on the fundamental ideas. 

A vector autoregression or VAR model is an atheoretical discrete-time stochastic model that captures collective dynamics of 
multiple time-series. This model allows us to generalize the analysis of co-movements from equal time to leading (and lagging) co-movements for all orders of lead (and lag). The literature on VAR model is very well established in the domain of time series econometrics (for a detailed textbook treatment, readers can consult \cite{hamilton1994time} and \cite{lutkepohl2005new}). In the present work, we utilize the VAR model to construct a dependency network out of multiple times series; e.g. to construct the dependency network across stock returns. Although the VAR model captures the collective dynamics of the nodes, this model
by itself cannot uniquely identify direction and magnitude of shock propagation. One needs to externally impose
more conditions which are known as {\it identification restrictions}.
In this paper, we propose that two particular sets of network filters in the form of {\it planner maximally filtered graphs} (PMFG henceforth; \cite{ASTE200520}, \cite{Tumminello_10}) and {\it partially correlation planar graph} (PCPG henceforth; \cite{kenett2010dominating}), provide identification restrictions that allow us to uniquely identify and estimate the model, and pin down unique paths of distress propagation.

The basic idea of both of the graph filtering techniques (PMFG and PCPG) depend on the concept of planar graph  (\cite{west1996introduction}) i.e., graphs that can be embedded in the plane with none of its edges crossing each other. 
Essentially, the PMFG and PCPG algorithms extract subgraphs out of the full graph with maximal information content. The method can be applied to a static, weighted graph with say $N$ nodes and $N(N-1)/2$ edges (for undirected graphs), $N(N-1)$ edges (for directed graphs). The filtered graph would have $3(N-2)$ number of edges (\cite{Tumminello_10} and \cite{kenett2010dominating}).
Some remarkable properties of the filtered planar graphs are: one, such graphs have closed loops and cliques of three or four nodes; and two, the graph remains connected and in particular, it contains the minimum spanning tree (MST henceforth). \cite{Tumminello_10} discussed that the filtered graph contains substantially more information than MST and at the same time, the loop and clique structures are preserved which are absent in MST. In the present context, we utilize these unique properties to characterize feedback loops in a connected network, that arises out of filtering a full information network characterizing an interlinked dynamical system.

The main intuition of our algorithm is as follows. The VAR estimation allows us to capture dependence of across many different time series and hence, allows us to create a dependency network. But it retains all possible signal of co-movements of all pairs of time series. With the help of the filtered networks, we can retain all the informative
edges, i.e. the ones that represent most important co-movements. Since as a by-product the filtered network has closed loops of three and four nodes, it allows us to study the effects of feedback loop in the dynamical set up. Thus by combining the properties of graph filtering which are useful for static graphs, with a dynamical system in the form of a VAR model, we can find out distress propagation as a dynamical response through the feedback loops that carry the maximal filtered information. We also show that since PCPG admits a directional network, it provides a more realistic and accurate description of distress propagation on empirical networks.

After developing the tools and techniques, we illustrate usage of the same through a set of applications. 
Generally speaking, the methodology can be applied to any co-evolving dynamical system in discrete time. Here we have chosen
two types of variables that evolve globally with substantial interdependence, viz. gross domestic product per capita and stock indices across the globe\footnote{Data source: Eikon database from Thompson Reuters; https://eikon.thomsonreuters.com/index.html.}. GDP per capita evolves slowly and we model its evolution with quarterly data.\footnote{We actually model the business cycle component. We will explain it later in more details (see Sec. \ref{subsec:shock_gdp_network}).} Stock indices are modeled at monthly frequency. These two different applications demonstrate the range of applicability of the proposed methodology. Due to generality of the proposed algorithm, we can model higher frequency time series without any change in the proposed algorithm.

We establish three points as the main contributions. One, VAR model identified with restriction implied by PCPG can be more accurately estimated than the same using restrictions implied by PMFG. Two, VAR-PCPG provides a very intuitive explanation of distress propagation due to its directional structure. Three, VAR-PCPG allows us to estimate impulse response functions through the feedback loops, thus illuminating the mechanism for diffusion of shocks in a complex network.  
This paper belongs to the newly emerging literature on network theory inspired time series econometrics (\cite{Barigozzi_17},
\cite{barigozzi2017generalized}, \cite{Diebold_17}, \cite{Diebold_book_15}). The applications on the economic and 
financial variables make it useful for studying shock spillover (\cite{Sharma_17}, \cite{corsi2018measuring})
on hierarchical networks (\cite{Tumminello_10}, \cite{di2010use}).
This paper is also related to the work presented in \cite{kumar2019ripples} which provided a mechanism for shock propagation by rank-ordering nodes in terms of centrality and the mechanism was agnostic towards local topology of the network. In this paper, the novel feature arises in terms of shock propagation through planar graphs and therefore, differences in local topology across the graph creates heterogeneity in propagation.

The plan of the paper is as follows. In Sec. \ref{sec:methodology}, we first describe the algorithm along with a step-by-step guide for implementation. In Sec. \ref{sec:data}, we describe the economic and financial data for application and we explain the spillover effects that can be characterized through the algorithm. Finally, Sec. \ref{sec:summary} summarizes and concludes the paper.

\vskip 1 cm


\section{Methodology: VAR estimation with planarity based graph filtration}
\label{sec:methodology}
Our proposed methodology uses vector autoregressive model (VAR) as the fundamental building block, along with planarity-based filtration techniques 
to uniquely identify and estimate the parameters of the model. Since the VAR model
provides a dynamic interpretation of multiple co-moving time-series, it is a wide range of applications in financial econometrics (\cite{Diebold_book_15}). However, the main difficulty arises in terms of estimation in two forms: one, a many-variable VAR model can be numerically challenging to estimate, and two, the sequence of shock propagation across time and entities needs to be identified by exogenously imposed restrictions.

Following the work of \cite{Mantegna1999}, many studies analyzes financial system as a weighted network with assets as its constituent nodes and weight of the edges as the strength of interaction between nodes. Drawing from this stream of literature, we argue that in the financial network, shock in one node propagate to other node through the connecting edge.  However, a complete and undirected graph/network of $N$ nodes has all possible $N(N-1)/2$ edges, which also include edges containing low information or statistically insignificant relationships. It is standard in network literature to rely on some filtration technique to retain only informative edges from the complete graph. In this study we rely on two planarity based filtration techniques, {\it Planar Maximally Filtered Graph} (PMFG henceforth) (\cite{ASTE200520}, \citep{Tumminello10421}) and {\it Partial Correlation Planar Graph} (PCPG henceforth) \citep{kenett2010dominating}. These filtration techniques retains only $3(N-2)$ edges by construction and generate connected subgrabh with maximal information of the underlying topological structure. As we detail below, such a connected subgraph yields required constraints for the identification of the VAR framework.

\subsection{Vector Autoregressive Models (VARs)}
 The VAR($p$) framework represents the evolution of $N$-dimensional time series $\bm{x}_t$ as a function of their lagged values of the order $1 \le \ldots \le p$ along with cross-dependence and a vector of error terms (\cite{lutkepohl2005new}, \cite{hamilton1994time}):
 \begin{equation}    \label{eqn:var}
 \begin{bmatrix} 
    x_{1t} \\ \vdots \\x_{Nt}
    \end{bmatrix}= \begin{bmatrix} 
    a^1_{11} & \dots & a^1_{1N} &  \\
    \vdots & \ddots & \vdots\\
    a^1_{N1} &        & a^1_{NN} 
    \end{bmatrix} \begin{bmatrix} 
    x_{1t-1} \\ \vdots \\x_{Nt-1}
    \end{bmatrix} + \ldots +\begin{bmatrix} 
    a^p_{11} & \dots & a^p_{1N} &  \\
    \vdots & \ddots & \vdots\\
    a^p_{N1} &        & a^p_{NN} 
    \end{bmatrix} \begin{bmatrix} 
    x_{1t-p} \\ \vdots \\x_{Nt-p}
    \end{bmatrix}+ \begin{bmatrix} 
    u_{1t} \\ \vdots \\u_{Nt}
    \end{bmatrix}
 \end{equation}
 or, in matrix form
\begin{equation*}\label{Eqn:VAR_short}
\bm{x}_t = A_1 \bm{x}_{t-1} + \ldots + A_p \bm{x}_{t-p} +\bm{u}_t, 
\end{equation*}
where $\bm{x}_t = (x_{1t}, \ldots, x_{Nt})$ denote $N$-dimensional time series, $A_k$ denotes $(N \times N)$ coefficient matrix with elements $a^k_{ij}$ where $k$ corresponds to the lag order, $i,j$ are row and column index respectively, $\bm{u}_t= (u_{1t}, \ldots, y_{Nt}) $ is a $N$-dimensional error vector with mean
$E(\bm{u}_t) = \bm{0}$ and covariance matrix $E(\bm{u}_t \bm{u}_t^\top) =
\Sigma_{\bm{u}}$. VAR($p$) models can also be equivalently written as Wold moving average form of the error terms by inverting the AR-polynomial in Eqn. \ref{eqn:var} (\cite{lutkepohl2005new}, \cite{hamilton1994time}) as: 
\begin{equation}
\label{eqn:wold}
\bm{x}_t = \Phi_0 \bm{u}_t + \Phi_1 \bm{u}_{t-1} + \Phi_2
\bm{u}_{t-2} + \ldots
\end{equation}
where  $\Phi_0 = I_K$ and $\Phi_s$ are $N \times N$ matrices that can be computed according to the following recursive relationship:
\begin{equation}
\label{eqn:wold_phi}
\Phi_s = \sum_{j=1}^s \Phi_{s-j} A_j \; \hbox{for} \; s = 1, 2, \ldots  
\end{equation}
Since right hand side of Eqn. \ref{eqn:wold_phi} only contains the error terms (shocks) and its lags, it allows us to identify the propagation of shock in the system. Element $\{\Phi_n\}_{i,j}$ is the response of $x_{i,t+n}$ to a unit impulse in $x_{j,t}$ with all other variables being held constant. This form of analysis of impulse response functions requires that a shock emanating from an epicenter affects only that epicenter at that time. In order to find the spillover the shock should not be affecting more than one nodes, in which case we will get an aggregate response which cannot be uniquely attributed to only one shock. In other words, we need to orthogonalize the system.  
To overcome this, \cite{SIM1980} proposed the idea of orthogonalizing the reduced form shocks using Cholesky decomposition of the covariance matrix of the error term
\begin{equation}
\Sigma_{\bm{u}} = PP'
\end{equation}
where, $P$ has a lower triangular form.
 $\bm{x}_t$ expressed in Eqn. \ref{eqn:wold} can easily be transformed into an alternative representation using $P$
 \begin{equation}
 \bm{x}_t = \Theta_0 \bm{\omega_t} + \Theta_1 \bm{\omega_{t-1}} + \Theta_2 \bm{\omega_{t-2}} + \ldots 
\end{equation}
where $\Theta_i = \Phi_i P$ and $\bm{\omega_t}= P^{-1} \bm{u}_t$.
Thus the resultant impulse response functions are dependent on the exact ordering of the variables. Because of the lower triangular structure of the matrix $P$, shocks propagate from top to bottom and not the vice versa.

Structural VAR (SVAR henceforth) models put additional restrictions in the form of $A$ and $B$ matrix on the reduced form VAR model described above to uniquely identify the propagation of shocks:
\begin{equation}
\label{Eqn:svar}
A \bm{x}_t = A_1 \bm{x}_{t-1} + \ldots + A_p\bm{x}_{t-p} + B \bm{\varepsilon}_t. 
\end{equation}
Since the number of estimable parameters in each matrix $A_j$ ($j=~1,~\ldots,~p$) above would be $N^2$, we use a parsimonius model retaining the dynamics in the following form (in our actual estimation process this is also chosen by the Bayesian information criteria as the optimal model):
\begin{equation}
\label{Eqn:svar_est}
A \bm{x}_t = A_1 \bm{x}_{t-1} + B \bm{\varepsilon}_t. 
\end{equation}

Empirical analysis rely on theoretical arguments to impose identifying restrictions. In the present context, we assume that shocks in the financial network propagate only through the informative edges constituting a connected subgraph. Filtered graph yields a sparse subgraph $G$ which provides the required restriction for the VAR identification. In particular, we use $B$ type of SVAR (see \cite{vars} for details of implementation) specification by setting $A$ as an identity matrix in Eqn. \ref{Eqn:svar_est}. The $B$-type SVAR specification requires additional $N(N-1)/2$ restrictions on the $B$ matrix for the identification. By restricting elements in the $B$ matrix corresponding to the non-connected edges to zero, we get the desired structure of the shock propagation.
As we will discuss below, planarity-based filtered graphs retain $3(N-2)$ edges. Therefore, the restrictions would boil down to estimating $3(N-2)$ number of elements in the $B$ matrix (twice of that in case of PMFG for symmetry reasons; we will elaborate on it below) while setting all other elements equal to zero. Interestingly, these restrictions are sufficient ($> \frac{N(N-1)}{2}$) for the identification for the VAR framework for all of our datasets. In particular, one can show it mathematically that for the identification restrictions to work, one needs a network of at least 13 edges (proof given in App. \ref{App:Iden}).

\subsection{Filtration based on graph planarity}

Based on the above discussion, we note that our objectives are two-folds. One, we need to impose restrictions on the $B$ matrix in Eqn. \ref{Eqn:VAR_short} to estimate the model ($A$ is set to identity matrix since equal time influence across nodes is absent in the present case). Two, the restrictions need to be such that the shock propagation mechanism can be modeled through the estimated process. We propose that a filtered planar graph is an ideal candidate satisfying both criteria. 

Before getting into the details, let us first discuss the objective. The goal is to clearly identify the paths of shock propagation across a network. One immediate choice would be minimum spanning tree (MST henceforth), which has often been used for visualization of financial networks. By its own definition, an MST retains $N-1$ edges such that the graph is still connected and the sum of the retained edges have the least weights.
However, that does not serve our purpose as the construction {\it filters the network too much} be removing all loops. Therefore, in order to study shock propagation through loops
we need a larger subgraph than MST; but it cannot be so large that it is not uniquely estimable.

We claim in this paper planarity-based filtered graphs achieve both the targets. It allows filtering the edges to the extent that the remaining subgraph is still connected and can be used for estimation of Eqn. $\ref{Eqn:VAR_short}$ with unique parameters. Also, by construction it retains all the loops of three and four nodes that makes it useful to actually study shock propagation in a complex fashion.  
Below, we provide the technical backgrounds on two graph filters based on planarity.

\subsubsection{Construction of Planar Maximally Filtered Graph}
\label{subsubsec:pmfg}

\begin{figure}[t]
	\centering
	\includegraphics[width=.5\linewidth]{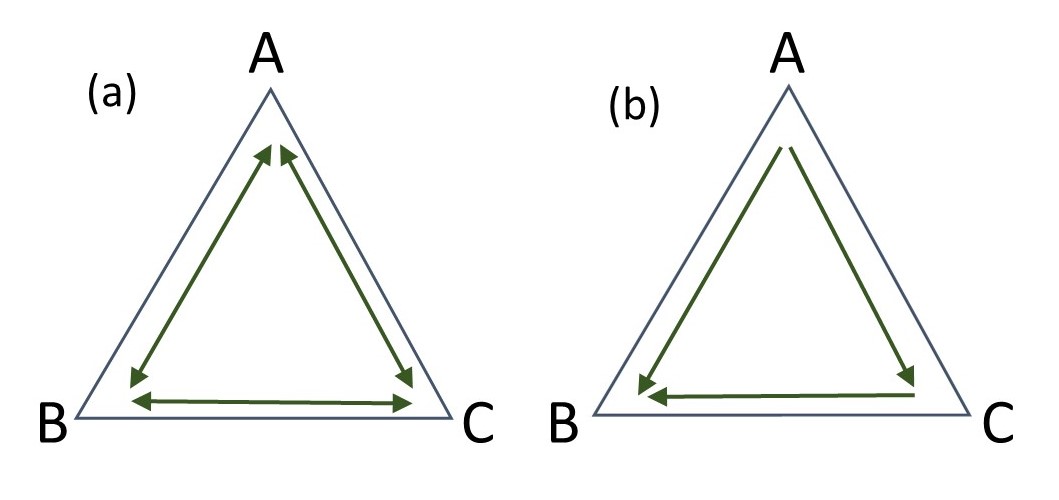}
	\caption{Two networks with loops. {\it Panel (a):} Undirected triad arising out of planar maximally filtered graphs (PMFG). {\it Panel (b):} A possible realization of directed triad arising out of partial correlation planar graph (PCPG). In the second network, the shock propagation would be asymmetric in both direction and magnitude.
    }
	\label{figure:network_dir_undir}
\end{figure}

As has been discussed above {\it Planar Maximally Filtered Graph} (PMFG) \citep{ASTE200520, Tumminello10421} retains loops and cliques of three and four nodes in the network. PMFG uses Pearson correlation ($C_{ij}=[E(x_i.x_j)-E(x_i)E(x_j)]/\sigma_i\sigma_j$ where $x_i$ and $x_j$ denote time series with means $E(x_i),~E(x_j)$ and standard deviations $\sigma_i,~\sigma_j$) matrix as an adjacency matrix for the complete graph. Pearson correlation is an aggregate measure of association between two variables $i$ and $j$, which also includes the influence of other variables on both $i$ and $j$. Further, adjacency matrix of the PMFG graph is symmetric by construction, suggesting identical influence of two variables on each other. In case of economic and financial networks, a more likely scenario would be asymmetric effects within a pair of nodes (the influence of large economies on smaller ones are typically much larger than the reverse). Visually, the PMFG retains closed loops like the one exhibited in {\it panel (a)} of Fig. \ref{figure:network_dir_undir}.  

Ideally, we would like to avoid these two features and create a network where edges would possess two features: one, the informative edges would retain the information after {\it controlling for} effects of all other nodes and two, the network should be weighted allowing for asymmetric impacts and responses across pairs of connected nodes. A newer construction allows us to precisely attain these objectives.

\subsubsection{Construction of Partial Correlation Planar Graph} 
\label{subsubsec:pcpg}

\cite{kenett2010dominating} provided an alternative by adapting PMFG in the directed/asymmetric setting. Using partial correlation, this methodology create an asymmetric adjacency matrix where the $\{i,j\}$-th element represents the measure of influence of variable 
$j$ on variable $i$ controlling for influences of all other variables. In order to calculate the partial correlation between a pair $\{i,~j\}$, let us imagine the third variable $k$ which also affects and is affected by both $i$ and $j$.
Then the interaction strength is defined as
\begin{equation}
PC_{ij|k}=\frac{C_{ij}-C_{ik}C_{jk}}{\sqrt{1-C_{ik}^2}\sqrt{1-C_{jk}^2}}.
\end{equation}
Given the partial correlation matrix, one can directly implement the planarity-based construction of the filtered graph following \cite{Tumminello10421}.
The outcome is a planar graph with filtered information that retains $3(N-2)$ edges along with directed and weighted structure, known as {\it Partial Correlation Planar Graph} or PCPG in short. Visually, the PCPG retains loops like the one exhibited in {\it panel (b)} of Fig. \ref{figure:network_dir_undir}. Here, node $A$ affects $B$ and $C$ and node $C$ affects $B$. Thus the spillover effects would be asymmetric.

We implement the VAR methodology on both PMFG and PCPG. We will discuss below that PCPG not only provides a more intuitive interpretation of shock propagation, but also has better properties that allows for more efficient statistical estimation of the VAR model.

\subsection{Estimating the SVAR model: Numerical optimization of the log-likelihood function}

The parameters of the SVAR models are estimated by optimizing the following log-likelihood function: 
\begin{equation}
\label{Eqn:loglik}
log~ \mathscr{L}( B) = - \frac{NT}{2}\ln(2\pi) -\frac{T}{2}\ln|B|^2  - \frac{T}{2}tr({B^{-1}}^\top B^{-1}\tilde{\Sigma}_u),
\end{equation}
where $N$ and $T$ denote the number and length of time series, $tr(.)$ denotes trace of a matrix, $\tilde{\Sigma}_u$ is the estimated reduced form covariance matrix of the error term (obtained from fitting Eqn. \ref{eqn:var} to $N$-dimensional time-varying data). Note that the matrix $B$ has a size of $N\times N$ and we conduct the estimation process by imposing zeros on all elements $B_{ij}$ for pairs $\{i,~j\}$
which have zero weights in the filtered planar graphs (both PMFG and PCPG).

The likelihood function is non-linear in nature and the corresponding parameter space is many dimensional, as number of free parameters in $N$ dimensional VAR framework is of $O(N^2)$. It suffers from  two well documented problem of convergence \citep{guo2016high}: the likelihood surface can be flat leading to non-existence of unique minima and possible convergence to a local minima. While it is beyond the scope of this study to make a claim regarding reaching global minima with complete certainty, we follow the practice in the literature and have implemented a series of different numerical optimization algorithms with may randomly chosen starting points to increase its chances for global convergence. In particular, we use Nelder-Mead \citep{Nelder} and BFGS \citep{fletcher2013practical}\footnote{We have also implemented Conjugate Gradient \citep{fletcher1964function} and Simulated Annealing \citep{belisle1992convergence} algorithms. However these two algorithms performed worse than Nelder-Mead and BFGS in the present application. Therefore, we do not report the results here.}. Nelder-Mead is a simplex based method and is shown to be robust but slow \citep{koshel2002enhancement}. However, there are some evidence of reduction in its efficiency as dimensionality increases \citep{han2006effect}. BFGS is a slope-based quasi-Newton method, which generally performs well for smooth convex functions.

We will describe the details of implementation model below in Sec. \ref{sec:data}. However, for the sake of completeness we lay out the overall features that we saw appears in the estimation process. We are presenting a summary of the results to help the reader to follow empirical viability of the two proposed numerical techniques. 

\begin{figure}[t]
      \centering
      \includegraphics[width=\linewidth,height=\textheight,keepaspectratio]{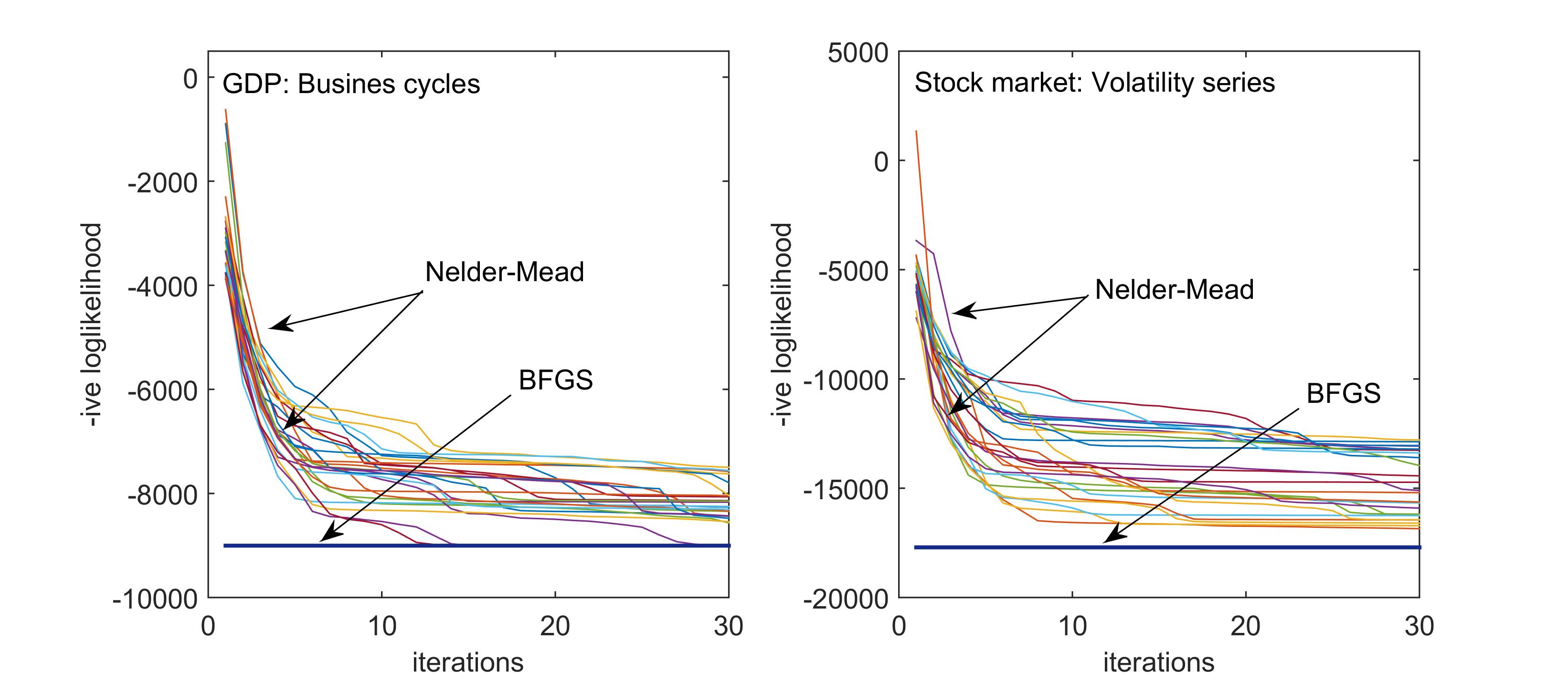}
      \caption{Likelihood convergence of the optimization for VAR estimation with two methods, viz. Nelder-Mead (downhill simplex; Convergence is shown for 25 different runs) and BFGS (hill climbing; only the best performing path is shown out of 25 different runs). {\it Left panel:} Likelihood optimization for GDP data.
      {\it Right panel:} Likelihood optimization for stock volatility data. Comparison of results from both the analysis exhibits a trade-off between efficiency of Nelder-Mead and BFGS algorithms. The best performing BFGS path is at least as good as all paths followed by the Nelder-Mead method; but on an average Nelder-Mead paths do show convergence although slowly whereas BFGS might get stuck in local optima (omitted due to very low accuracy).}
     \label{figure:likelihood}
  \end{figure}

As both of these methods can converge to any possible local minima, we adapt a multistart strategy to search for global minima. For each of the likelihood optimization exercises, we perform 25 independent runs of optimization algorithm starting with different randomly chosen sets of parameters. Relevant for our case, Nelder-Mead algorithm also shows tendency of inappropriate termination \citep{Russell} even before reaching a local minima. To mitigate this, for every run we perform 30 iterations of the algorithm, using estimated  parameters of the previous iteration as a starting value for the next iteration (improvement in likelihood becomes marginal with larger number of iterations and according to our numerical analysis, the improvement  is negligible beyond 30 iterations; however, since it is a numerical search problem, a well recognized norm is that one cannot really claim with complete certainty that the global optima has been reached). In total, we run $750~(~=~25\times 30)$ rounds of optimization for a given likelihood function with each method. Fig. \ref{figure:likelihood} displays the evolution of negative log likelihood for the whole set of runs from the starting points to 30 iterations. For both GDP and stock market volatility series, the best performing BFGS provides us lower negative log likelihood values. 

However, there is a trade-off. In almost all runs, Nelder-Mead shows a slow convergence (Fig. \ref{figure:likelihood}) whereas other BFGS runs under-perform and gets stuck in local optima which are substantially worse than the Nelder-Mead solutions (not shown here).
Therefore, we conclude that in the present application, Nelder-Mead is slow but robust whereas the best run of BFGS (out of many) can potentially outperform Nelder-Mead, but it is non-robust since its convergence depends on the starting point.

%


\subsection{Implementation of the Proposed Algorithm}
\label{subsec:methodology}

For completeness, here we provide a detailed step-by-step description of the algorithm we propose in the paper, using both types of filtering methods.

\subsubsection{Shock propagation: SVAR with PMFG filtered network}
\begin{enumerate}
\item{\textbf{Extract subgraph $G_{PMFG}$ through PMFG filter} 
\begin{enumerate}
\item {\bf Correlation matrix:} Given $N$ time series, calculate $N \times N$ Pearson correlation matrix $C$, such that $C_{ij}=[E(x_i.x_j)-E(x_i)E(x_j)]/\sigma_i\sigma_j$ where $x_i$ and $x_j$ denote time series with means $E(x_i),~E(x_j)$ and standard deviations $\sigma_i,~\sigma_j$.
\item {\bf Ordered list:} Create an ordered list $S_{ord}$ of all the $\frac{N(N-1)}{2}$ elements of lower triangular component of $C$ in decreasing order.
\item {\bf Adjacency matrix construction:} Following the ordered list $S_{ord}$ from the top, add an edge between nodes $i$ and $j$ if and only if the graph is still planar after adding the edge. Else, remove the added edge and repeat the same exercise for the next element in the ordered list.
\item {\bf Network construction:} Transform lower triangular matrix into a symmetric PMFG matrix by adding its transpose.
\end{enumerate}
}

\item{\textbf{Estimate Structural VAR model} 
\begin{enumerate}
\item {\bf Identification restrictions:} $\forall$ $i \leq N$, $j\leq N$ , $i \neq j$ and  $G_{PMFG}(i,j)=0$, restrict $B(i,j)=0$.
\item {\bf Estimation:} Estimate $B$-type SVAR specification with the restricted $B$ matrix using maximum likelihood estimator.
\end{enumerate}
}

\item{\textbf{Estimate impulse response function} 
\begin{enumerate}
\item Estimate the impulse response function from the fitted SVAR model. This will provide the exact path of the shock propagation in the PMFG network.
\end{enumerate}}

\end{enumerate}

\subsubsection{Shock propagation: SVAR with PCPG filtered network}
\begin{enumerate}
\item{\textbf{Extract subgraph $G_{PCPG}$ through PCPG filter}}
\begin{enumerate}
\item {\bf Partial correlation matrix:} Given $N$ time series, calculate $N \times N$ Pearson correlation matrix $C$ and $N \times N \times N$ partial correlation matrix $PC$.
The influence of variable j on the correlation $C_{ik}$ is calculated as
\begin{equation}
d(i,k|j) = C_{ik}- PC_{ik|j}
\end{equation}
\item {Influence matrix:} The total influence of the variable j on i $D(i,j)$ is the average of  $d(i,k|j)$ over ks.
\begin{equation}
D(i,j)= \frac{1}{N-1} \sum_{k\neq j}^{N-1}d(i,k|j)
\end{equation}

\item {\bf Ordered list:} Create an ordered list $S_{ord}$ of top $N(N-1)/2$ elements of $D$ in decreasing order, such that $\forall$ $i \leq N$, $j\leq N$ if $D(i,j) \in S_{ord} \implies D(j,i) \notin S_{ord} $.
\item {\bf Adjacency matrix construction:} Following the ordered list $S_{ord}$ from the top, add an edge between
nodes $i$ and $j$ if and only if the graph is still planar after adding the edge. Else, remove the added edge and repeat the same exercise for the next element in the ordered list.
\end{enumerate}
\item {\bf SVAR and shock propagation:} Step 2 and Step 3 of this method is similar to the method discussed above using PMFG filtration. 
\end{enumerate}

\section{Applications of the Algorithm: Shock Propagation on Economic and Financial Networks}
\label{sec:data}

In this section, we apply the algorithm we proposed above by combining structural VAR with graph planarity-based identification restrictions, to real world networks.
In order to demonstrate the usefulness and to highlight important features of the methodology, we have chosen two specific cases.
First, we study shock spillover across the global economy in terms of economic fluctuations. For this purpose, we analyze the {\it business cycles} of the G-20 countries
and analyze the shock propagation on the corresponding fluctuation network.
Second, we analyze shock spillover across the global economy 
in terms of financial volatility.
For this purpose, we study the financial market volatility of G-20 countries and characterize the shock propagation on the
corresponding financial volatility networks.
In table \ref{table:data} we provide details of these two sets of data along with their frequencies and the source of data.

\begin{table}[tbp]
  \centering
  \caption{Data description: GDP and stock markets}
   \label{table:data}
    \begin{tabular}{p{2cm}p{3.5cm}p{2cm}p{2cm}p{4cm}}
    \toprule
    Variable & Dataset & Frequency & Period & Country \\
    \midrule
    GDP & Cyclical component extracted using HP filter of GDP series & Quarterly & Q1 1980 - Q4 2018 & ARG, AUS, BRA, CAN, CHN, FRA, GER, IND, INDO, ITA, JPN, MEX, SA, SKOR, TUR, UK, US (G-20 countries)  \\
    Stock return volatility & Latent volatility series of stock indices & Monthly & Jan 2000 - Dec 2018 & ARG, AUS, BRA, CAN, CHN, FRA, GER, IND, INDO, ITA, JPN, MEX, RUS, SA, SKOR, TUR, UK, US (G-20 countries)\\
    \bottomrule
    \end{tabular}%
\\
{~~~~~\small Source: Thomson Reuters Eikon database; https://eikon.thomsonreuters.com/index.html}\hfill
  \label{tab:addlabel}%
\end{table}%

\subsection{Shock Spillover on Business Cycle Network}
\label{subsec:shock_gdp_network}

First, we analyze shock propagation in the economic network of G-20 countries. We have 156 (4 observations each year, for 39 years) quarterly observations for 17 countries from Q1-1980 to Q4-2018. 
Since GDP data contains both trend and cyclical components, it is a standard practice
to decompose the series in two parts. In the following we employ Hodrick-Prescott filter
(\cite{Hodrick}) to carry out the decomposition\footnote{HP filter is a very well known toolkit in dynamic macroeconomic literature that allows us to decompose time series into a trend and a cyclical components by suitably tuning a penalty parameter. Interested readers can refer to (\cite{dejong2011structural}) for a textbook exposition of the technique.} and following the literature, we extract the cyclical component which conveys information about {\it business cycles}.
Then we construct the business cycle network and estimate a structural VAR model with planarity-based restrictions as have been described in details in Sec. \ref{subsec:methodology}. 
The estimated impulse response function allow us to analyze shock propagation using both the filtration methods. 

\begin{figure}[t]
      \centering
      \includegraphics[width=0.56\linewidth]{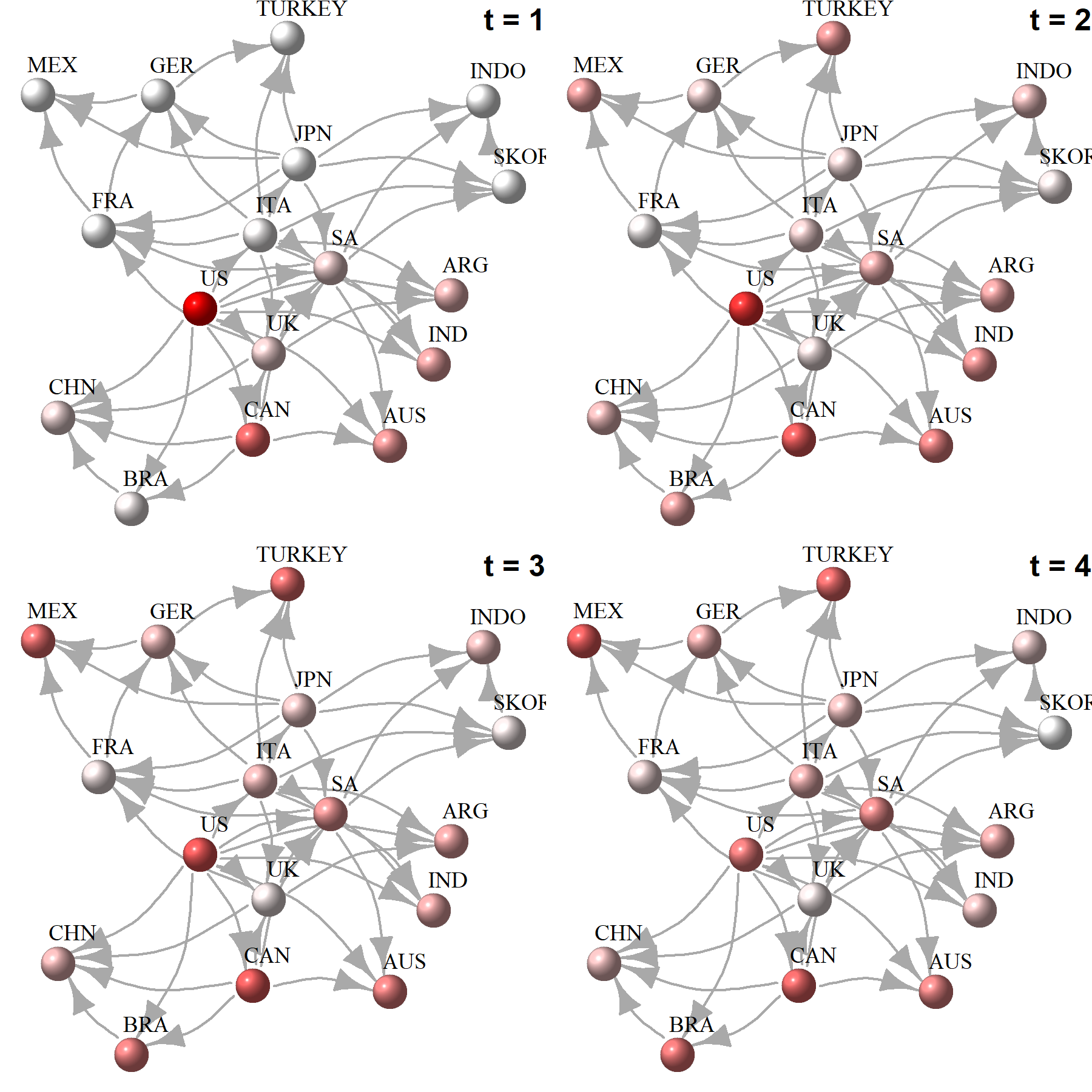}
      \caption{PCPG-SVAR estimation: shock propagation on G-20 GDP fluctuation network. US is chosen as the epicenter of shocks. We plot the impulse responses with respect to an unit shock to the US business cycles, estimated through the PCPG-SVAR model. The shock permeates through the network as well as diffuses over time as is indicated by the impulse response functions. The analysis above suggests that the response to the shock is most prominent in Argentina, Germany, Japan, UK, Canada and Australia (intensity of color indicates the magnitude of the response) whereas countries like India, China, Indonesia and South Korea are largely unaffected even after $t$ = 4 time points (equivalent to a year since the estimation carried on quarterly data).}
      \label{figure:PCPG_GDP}
  \end{figure}

As an empirical demonstration, we analyze how shocks from USA to spill over to other countries. The results are presented in 
Fig. \ref{figure:PCPG_GDP} using restrictions implied by PCPG. 
In App. \ref{subsec:shock_gdp_network}, we provide the results obtained from the PMFG restrictions as well (Fig. \ref{figure:PMFG_GDP_SI} left panel for business cycle network).
We notice that the shock spillover estimation through PCPG restrictions are much more intuitive than the PMFG restrictions on two counts. First, the SVAR estimation with PCPG correctly identifies the shock propagation with substantial effects of the shock on itself, showing persistence. This is fairly consistent with other empirical documentation of economic shock spillover. In comparison, in case of PMFG-based results, we find that the direct effect appears more pronounced in Argentina with very little impact on USA although the epicenter of the shock is USA. This is inconsistent with empirically documented persistence of GDP fluctuations at the country-level. Second, the shock spills over to the cluster of developed countries and the effects are less prominent in the developing countries like India, Indonesia, South Korea etc whereas it is more prominent for developing countries that are geographically close to USA, like Mexico, Argentina etc.

\subsection{Shock Spillover on Stock Index Network}
\label{subsec:shock_stock_network}

Next, we analyze the volatility shock spillover in the global financial market. In particular, we study stock markets of G-20 countries. Using monthly returns of major stock indices of 18 countries (data not available for two countries) for the period Jan. 2000- Dec. 2018 (table \ref{table:data})), we extract latent volatility series by fitting GARCH(1,1) model \citep{BOLLERSLEV1986307}. Generalized autoregressive conditional heteroskedasticity (GARCH henceforth) is a popular methodology to extract latent volatility of  financial return series. A standard GARCH$(p,q)$ framework is represented as: 
\begin{eqnarray}
r_t    &=&\sigma_t\epsilon_t, \nonumber \\
\sigma_t^2&=& \bar{c}+ \sum_{i=1}^{p}\alpha_ir^2_{t-i} +\sum_{j=1}^{q}\beta_j\sigma^2_{t-j},
\end{eqnarray}
where $r_t$ is the return series and $\sigma_t$ is latent volatility series which is unobservable in real data. The main goal of GARCH modeling is to estimate the time series of latent volatility $\{\hat{\sigma}_t\}$. In principle, one can utilize more complex toolkits for measuring latent volatlity. But GARCH has become the most standard and easy-to-implement model. Since our focus in simply to get an estimate of latent volatility, we stick to the choice of GARCH(1,1) for its simplicity.
 
\begin{figure}[t]
      \centering
      \includegraphics[width=0.56\linewidth]{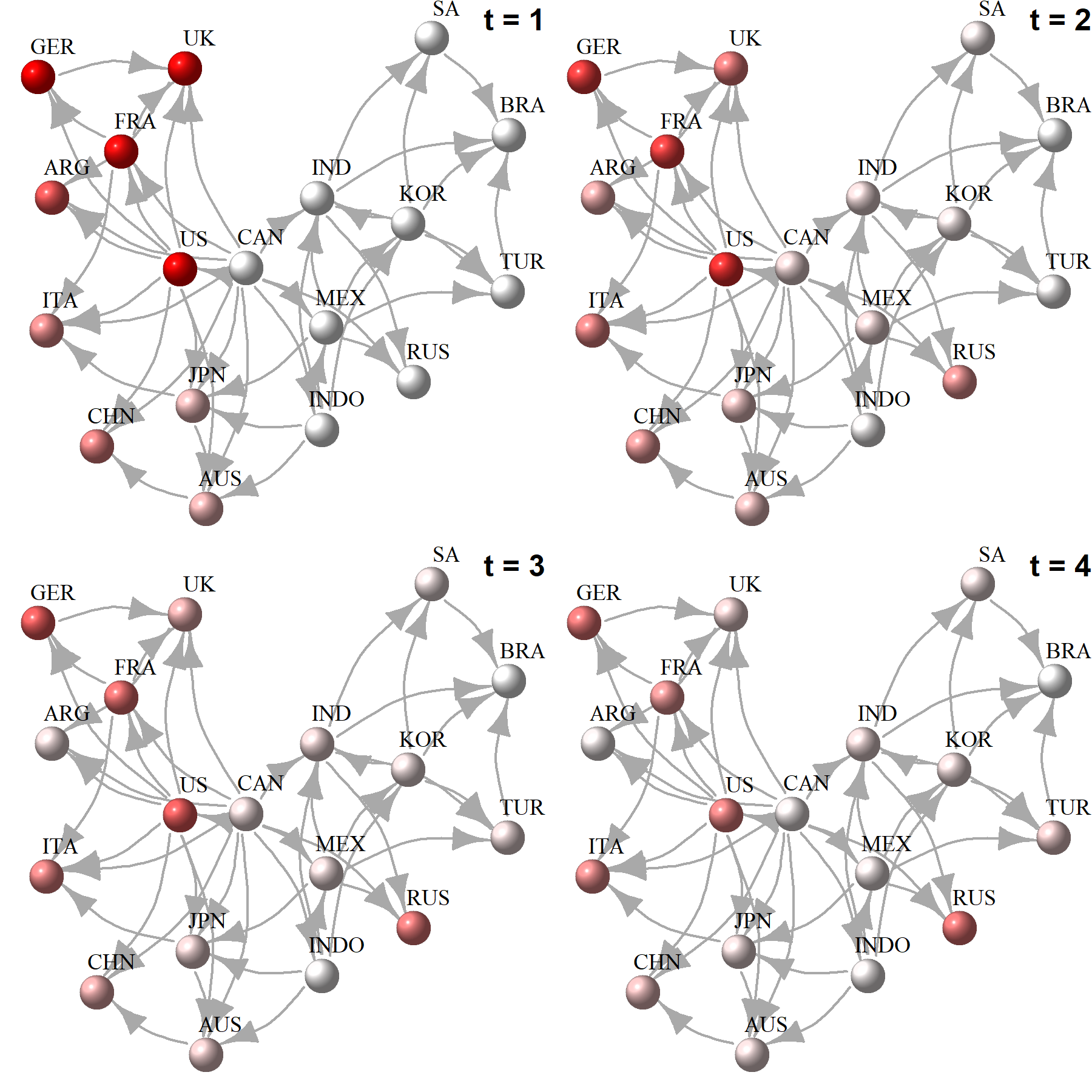}
      \caption{PCPG-SVAR estimation: shock propagation on G-20 stock index network with US at the epicenter. Estimation procedure and specification same as in Fig. \ref{figure:PCPG_GDP}. The analysis above suggests that the response to the volatility shock to S\&P 500 index is most prominent in the stock markets of Argentina, Germany, UK, Canada and France (intensity of color indicates the magnitude of the response) whereas countries like India, Indonesia, Brazil, Turkey and South Korea are largely unaffected even after $t$ = 4 time points (equivalent to a year since the estimation carried on quarterly data).}
      \label{figure:PCPG_SI}
  \end{figure}

Fig. \ref{figure:PCPG_SI} (Fig. \ref{figure:PMFG_GDP_SI} right panel in App. \ref{App:pmfg_figure}) represents how volatility shock from US propagates in the stock market G-20 countries using PCPG-SVAR (PMFG-SVAR). We note that PCPG-SVAR estimation exhibits 
direct spillover to other developed countries including United Kingdom, France, Germany along with some influence on Argentina, Japan and China. We see that after 4 time-points (equivalent to 4 months), the shock permeates and becomes muted in the affected countries. It is also noteworthy that countries like Indian South Korea, South Africa, Turkey were barely affected and show very muted response overall.

\section{Parsimony vs efficiency: A comparison between PCPG-SVAR and PMFG-SVAR}

In this section, we discuss relative merits and demerits of two different planarity-based restrictions for the estimation
of the SVAR model. We evaluate the methods in three dimensions, viz. applicability, parsimony and uniqueness. Our empirical exercises suggest that all three are inter-related and PCPG-SVAR dominates over PMFG-SVAR in all three counts.

First, we discuss applicability of the algorithms and suitability for general networks.
We demonstrate in App. \ref{App:Iden}, PMFG provides sufficient restrictions for the identification of the SVAR model only for $N\leq 2$ (we can ignore this condition for any meaningful network structure) or $N \geq 11$. On the other hand, PCPG provides sufficient restrictions for the VAR identification for all the integer values of $N\ge 4$. 
Thus PCPG-SVAR is generally speaking less restrictive in application than PMFG-SVAR.

Parsimony in the present context refers to the idea of providing desired level of explanation with fewer estimated parameters. PCPG-SVAR by its structure imposes more restrictions ($N(N-1)/2-3(N-2)$) with fewer free parameters ($3(N-2)$) whereas PMFG-SVAR imposes lesser restrictions ($N^2-2\times 3(N-2)$) with twice the number of free parameters ($2\times 3(N-2)$). Thus PCPG-SVAR is more parsimonius. 

Relatedly, a parsimonius model also helps to overcome the curse of dimensionality \citep{krolzig2001general} in the VAR based framework. Our empirical exercises suggest that PMFG-SVAR estimation results in flat likelihood surface and multiple local minima (results not reported in the manuscript) whereas numerical results show that PCPG-SVAR provides a smooth likelihood surface with enough curvature to carry out the optimization exercise.

Finally, we close the discussion by pointing to the fact that one can potentially argue that minimum spanning tree (MST) is in principle even more parsimonius than PCPG and therefore, should provide even better estimation properties. So a question might arise as to why are we arguing in favor of PCPG-SVAR rather than an equivalent of MST-SVAR?
It is interesting to note that, While MST can provide the most parsimonious structures to the  SVAR specification with $2(N-1)+N$ free parameters, it does not provide us the flexibility to explore possible feedback in the network. On the other hand, PCPG albeit less parsimonious than MST, retains sufficient number of edges that produces loops in the network and allow us to explore feedback effects. Additionally PCPG being the only filtering method among these three are based on directed network, allows us to capture asymmetric interaction between the nodes.
Thus we argue that PCPG-SVAR provides an excellent balance between MST and PMFG by retaining important topological properties of a network and provides enough degrees of freedom to estimate an SVAR model.

\section{Summary}
\label{sec:summary}

Real world networks are characterized by feedback loops, cascading failure and heterogeneous responses to external and internal shocks. Much of the applications of  network theory is dependent on credible modeling of distress propagation. More generally, recent research is focused on statistical estimation of networks from large scale real world data to model the link between failure at a node level and the corresponding macro-level response on the network as a whole.

In this paper, we provide an algorithm by combining tools from network filtering, along with econometric estimation of dynamic models to capture the same link between micro-level shock and macro-level repercussions. Specifically, we propose a structural vector autoregression model with identification criteria obtained from the topological properties of co-movement networks of dynamic variables.
We show that planarity-based restrictions on directed graphs provide an unique balance between applicability, efficiency and parsimony.
We apply the proposed algorithm to two major global networks, viz. economic fluctuations across countries and financial fluctuations across countries. The results clearly delineate the impact of shock diffusion, clustering of countries based on dynamic impacts and the importance of relative positions of the countries in the networks in terms of distress propagation.

The present work belongs to the stream of literature on systemic risk in large scale networks, especially those of economic (\cite{Acemoglu_16}) and financial nature (\cite{hue2019measuring}).
As for practical importance, it provides an avenue to feed into risk management in intertwined asset markets with nonlinear dependence
(\cite{bouchaud2003theory}). In the larger scheme of things, the algorithm presented in the paper would be relevant for understanding general features of cascading failures in complex economic as well as physical systems (\cite{duan2019universal}).

\newpage
\bibliographystyle{plainnat}

\newpage

\appendix

\section{Estimation of SVAR through planarity-based restrictions}
\label{App:Iden}

$B$-type of SVAR model of $N$ dimensional time series requires minimum $N(N-1)/2$ additional restrictions on $B$ matrix. Note that PMFG has $3(N-2)$ non-zero undirected edges by construction, which results in $2\times 3(N-2) $ non-zero elements in symmetric adjacency matrix. Further, we also need  diagonal elements ($N$) of the $B$ matrix unrestricted 
to capture persistence of shocks. Hence, total number  of restrictions implied by PMFG filtering on $B$ matrix is $N^2- 2\times3(N-2)-N $. Hence, for the identification of model,
\begin{equation}
N^2- 2\times3(N-2)-N \geq N(N-1)/2, 
\end{equation}
which implies
\begin{equation}
N\leq 2 ~~~\text{or} ~~~N \geq 11.
\end{equation}
We note that $N$ = 1 or 2 would lead to a trivial network. Therefore, the condition $N\ge 11$ is more useful for real world networks.

Similarly, asymmetric PCPG with $3(N-2)$ non-zero edges results in $3(N-2)+N$ unrestricted parameters. Therefore, we need the following inequality for identification:
\begin{equation}
N^2- 3(N-2)-N \geq N(N-1)/2, 
\end{equation}
which implies
\begin{equation}
N\leq 3 ~~~\text{or} ~~~N \geq 4.
\end{equation}
So clearly PCPG-SVAR is estimable for a larger class of networks with number of nodes greater than or equal to 4.

\section{Shock Spillover through SVAR-PMFG}
\label{App:pmfg_figure}

\begin{figure}
	\centering
	\begin{subfigure}{.48\textwidth}
		\centering
	\includegraphics[width=0.99\linewidth]{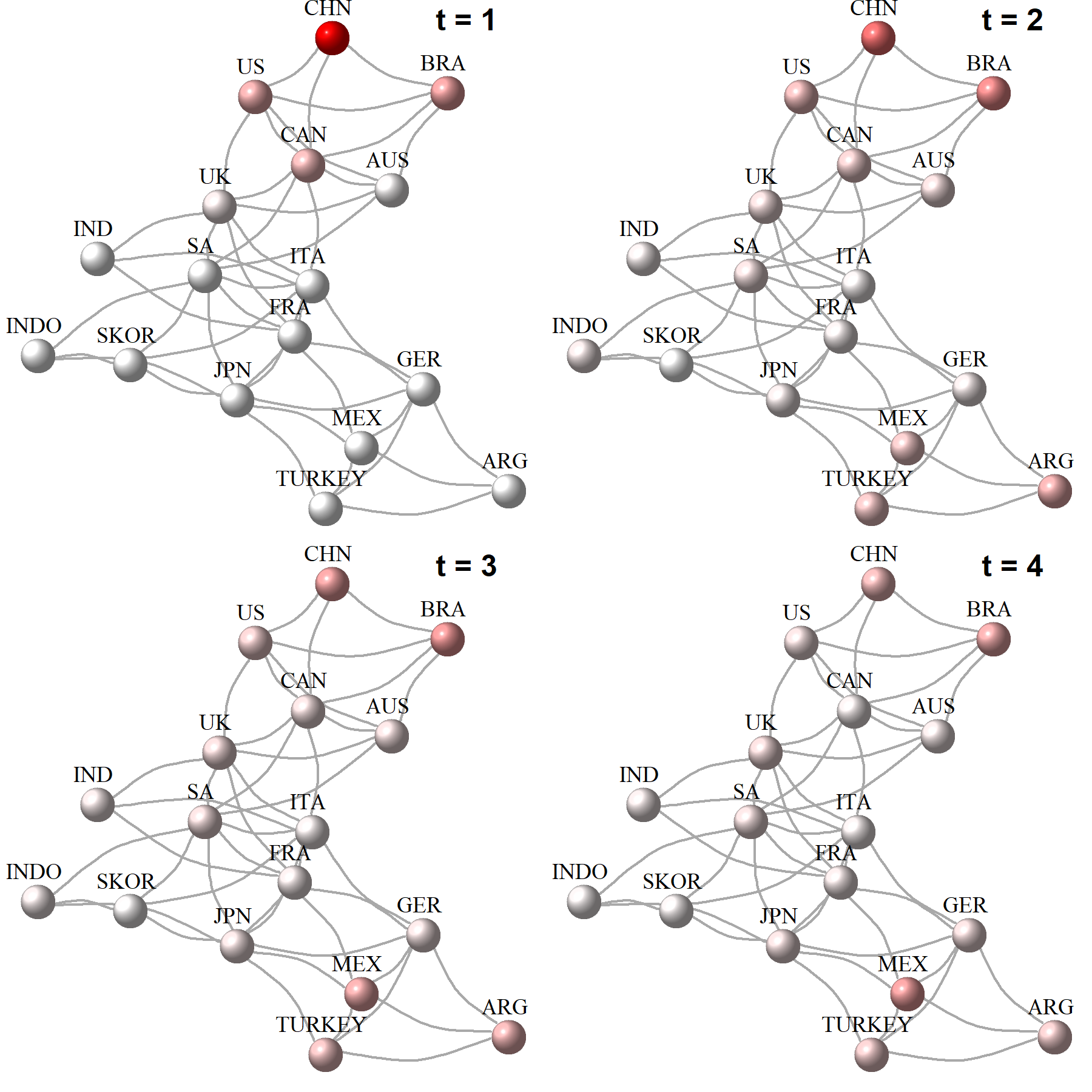}
\end{subfigure}
\begin{subfigure}{.48\textwidth}
	\centering
	\includegraphics[width=0.99\linewidth]{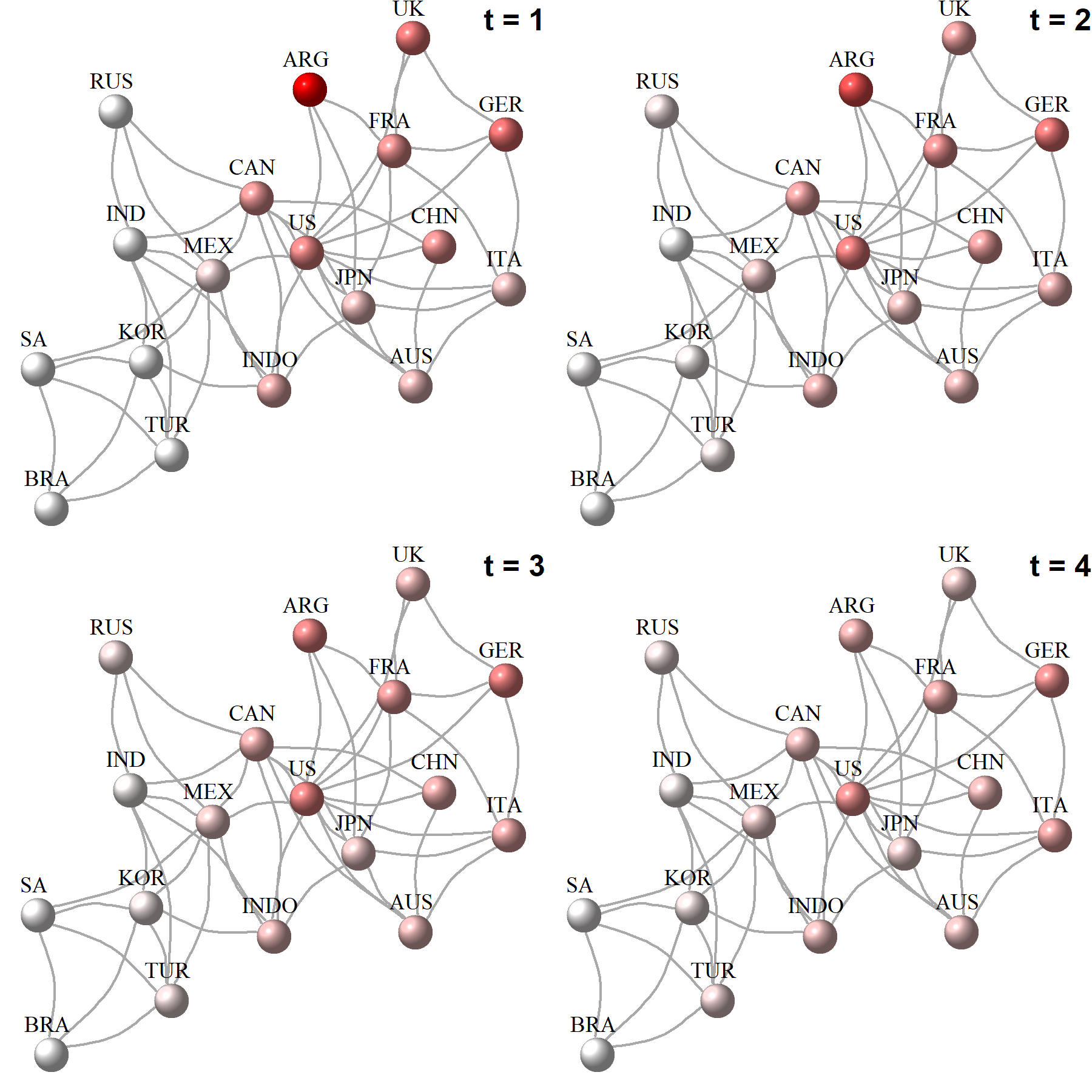}
\end{subfigure}
	\caption{Shock propagation on G-20 economic and financial networks. US is chosen as the epicenter of shocks in both cases. {\it Left panel:} PMFG-SVAR GDP growth network: We plot the impulse responses with respect to an unit shock to the US business cycle, estimated through the PMFG-SVAR model, and the diffusion of the shocks at $t$ = 1, 2, 3 and 4 quarters. {\it Right panel:} We plot the impulse responses with respect to an unit shock to the US stock index (S\&P 500) estimated through the PMFG-SVAR model, and the diffusion of the shocks at $t$ = 1, 2, 3 and 4 months.}
	\label{figure:PMFG_GDP_SI}
\end{figure}

Here we provide the figures (Fig. \ref{figure:PMFG_GDP_SI}) in two panels describing the shock propagation when the structural VAR model is estimated with restrictions implied by PMFG.

\end{document}